\documentclass[12pt]{article}

\usepackage{epsfig}
\usepackage{graphicx}
\usepackage[figuresright]{rotating}

\usepackage{latexsym}
\usepackage{authblk}

\usepackage{amsmath}
\usepackage{amssymb}
\usepackage{amsthm}
\usepackage{appendix}
\usepackage{cite}
\usepackage{indentfirst}
\usepackage[colorlinks,
            linkcolor=red,
            anchorcolor=green,
            citecolor=blue
            ]{hyperref}

\newcommand{\bea}{\begin{eqnarray}}
\newcommand{\eea}{\end{eqnarray}}
\newcommand{\bean}{\begin{eqnarray*}}
\newcommand{\eean}{\end{eqnarray*}}

\newcommand{\be}{\begin{eqnarray}}
\newcommand{\ee}{\end{eqnarray}}

\numberwithin{equation}{section}

\def\abs#1{\left| #1\right|}
\def\braket#1{\left\langle #1 \right\rangle}
\def\bra#1{\left\langle #1\right|}
\def\ket#1{\left| #1\right\rangle}
\def\bket#1{\left| #1\right]}
\def\gb #1{ \left\langle #1 \right]}
\def\tgb #1{ \left[ #1 \right\rangle}

\def\IC{\mathbb{C}}
\def\IP{\mathbb{P}}

\def\W #1{\widetilde{#1}}

\def\abs#1{\left| #1\right|}
\def\braket#1{\left\langle #1 \right\rangle}
\def\bra#1{\left\langle #1\right|}

\def\ket#1{\left| #1\right\rangle}

\def\gb #1{ \left\langle #1 \right]}
\def\tgb #1{ \left[ #1 \right\rangle}

\def\Label#1{\label{#1}}%

\def\bbar#1{ \overline #1}

\def\la{\lambda}
\def\tl{\tilde\lambda}

\def\eps{\epsilon}

\def\vev#1{\left\langle #1 \right\rangle}

\begin{document}
\baselineskip=18pt

\setcounter{footnote}{0}
\setcounter{figure}{0}
\setcounter{table}{0}

\title{Equivalence of Coefficients Extraction of One-loop Master Integrals
}

%
%
%
%
%
\author[1]{Yang An}
\author[2]{Zi-ang Hu}
\author[2]{Zhongjie Huang}
\author[1]{Yi Li \footnote{The corresponding author}}
\author[1]{Xiang Lv}
\affil[1]{Zhejiang Institute of Modern Physics, Department of Physics, Zhejiang University}
\affil[2]{Department of Physics, Zhejiang University}
\renewcommand*{\Affilfont}{\small\it} 
\renewcommand\Authands{ and } 
\date{} 
\maketitle
\begin{abstract}

Now there are many different methods to do the PV-reduction for the
one loop amplitudes. Two of them are unitarity cut method and
generalized unitarity cut method. In this short paper, we present an
explicit connection of these two methods, especially how the
extractions of triangle and bubble coefficients are equivalent to
each other.

\end{abstract}

\thispagestyle{empty}
\newpage
\setcounter{page}{1}
\tableofcontents
\section{Introduction}

Experiment of high energy physics such as LHC requires calculations of  cross section of  processes involving multiple particles. Precise theoretical predictions of these processes need calculations of Feynman diagrams at one-loop level and beyond. However, loop calculation is tedious and very inefficient using the standard method.
In recent years new methods are developed in order to prompt the calculation and avoid laborious work.
Currently, one-loop calculation has been considered as a solved problem and the focus is the higher loop calculations.

For loop calculation, the main approach is the Passarino-Veltman
(PV) reduction method \cite{Passarino:1978jh} . The reduction can be
divided into two categories: the reduction at the integrand level
and the reduction at the integral level. For one-loop case, the
efficient integrand level reduction is introduced by Ossola,
Papadopoulos and Pittau in \cite{Ossola:2006us}. For integral level
reduction, unitarity cut method and the generalized unitarity cut
method are two main used methods now.

Unitarity cut method was  introduced in
\cite{Bern:1993kr,Bern:1994cg}. The main idea is that since we know
the expansion
\bea {\cal A}^{1-loop}=\sum_i C_i I_i\eea
where  the one loop master integrals are  a set of scalar integrals, which is defined as
\footnote{In (\ref{master-def}) the $K_i$'s are sums of external momenta, which are strictly four-dimensional.
To  regularize the divergence, the loop integral is carried out
in $D=4-2\epsilon$ to regularize the divergence. For four-dimensional spacetime,  the range of $n$ is from $1$ to $4$ if considering only to $\eps^0$ order, or from $1$ to $5$ if all order of $\epsilon$ involves is needed.}
\bea I_n = (-1)^{n+1} i (4\pi)^{\frac{D}{2}} \int
\frac{d^{D}\ell}{(2\pi)^{D}}
\frac{1}{(\ell^2-m_1^2)((\ell-K_1)^2-m_2^2)((\ell-K_1-K_2)^2-m_3^2)\cdots
((\ell+K_n)^2-m_n^2)} . \Label{master-def} \eea
If we take the imaginary part of a given branch at both sides, we
will  have
\bea {\rm Im}{\cal A}^{1-loop}=\sum_i C_i {\rm Im} I_i~~~~\label{eq:mi}\eea
Thus if we can calculate the imaginary part easily, we can extract the master coefficients by comparing both sides. By Cutkosky rules \cite{Cutkosky:1960sp}, the calculation of imaginary part is doing following phase space integration
\bea
 \Delta A^{\rm 1-loop} \equiv \int d\mu~~ A^{\rm tree}_{\rm Left} ~\times~
 A^{\rm tree}_{\rm Right},
\Label{cutdef}
\eea
where the Lorentz-invariant phase space (LIPS) measure is defined by
\bea
d\mu = d^4{\ell_1}~ d^4{\ell_2}~ \delta^{(4)}(\ell_1+\ell_2 -
K)~ \delta^{(+)}(\ell_1^2)~\delta^{(+)}(\ell_2^2).
\Label{lips}
\eea
Here, the superscript $(+)$ on the delta functions for the cut propagators
denotes the choice of a positive-energy solution. Although the integration has been
simplified to two dimension, carrying it out is still a difficult task. The breakthrough
comes after the realizing that by holomorphic anomaly, such a two dimensional phase space integration
can be translated to read out the residue of corresponding poles \cite{Witten:2003nn,Cachazo:2004kj,Cachazo:2004zb,Britto:2005ha}. Using this technique,
analytic expressions of coefficients of one-loop master integrals have been given in series papers
\cite{Anastasiou:2006gt,Anastasiou:2006jv,Britto:2007tt,Britto:2008sw}.

Inspired by the double cut for the imaginary part, multiple cuts
have been also proposed in \cite{eden2002analytic,Bern:1992em} .
Especially, in \cite{Britto:2004nc} it has been shown that putting
four propagators on-shell, one can read out the coefficient of boxes
as the multiplication of four on-shell tree level amplitudes at the
four corners. This generalized unitarity cut method has been further
developed in\cite{Forde:2007mi,ArkaniHamed:2008gz}  .

Both methods, i.e., the unitarity cut method and the generalized
unitarity cut method, have solved the one loop integral level
reduction completely. However, the connection between these two
methods has still not been clearly demonstrated. It is our purpose
in this short paper to reveal the equivalence of these two methods.

The plan of the paper is following.  In section two we have reviewed
the two methods and in section three we will present our proof of
the equivalence of these two methods.

\section{Review of unitarity cut method and generalized unitarity cut method}\label{sec:um}

In this section, we will review both methods to establish the basis
for our investigation. In the first subsection, we will briefly
review the unitarty cut method and write down the major formula. We
will review the generalized unitarity cut method in the second
subsection.

\subsection{Review of unitarity cut method}

The unitary property of S-matrix means $S^\dagger S=1$. Writing  $S=1+i T$, we have $2 {\rm Im ~} T = T^\dagger T$, which is the familiar optical theorem. Expanding this equation by the order of coupling constant, we see that the imaginary part of the one-loop amplitude is related to a product of two on-shell tree-level amplitudes.
This imaginary part should be viewed more generally as a discontinuity across the branch cut singularity of the amplitude---in a kinematic configuration where one kinematic invariant momentum, say $K^2$, is positive, while all others are negative.  This condition isolates the momentum channel $K$ of our interest; $K$ is the sum of some of the external momenta.

As we have mentioned in the introduction, for one-loop amplitude the  Cutkosky rules  gives the integration  (\ref{cutdef}). Now we discuss how to carry out the phase space integration.
Since we are trying to compare with the method given in \cite{ArkaniHamed:2008gz},  we will focus on massless theory in pure 4D, thus we rewrite it as
\bea A^{\rm 1-loop}= - i (4\pi)^{2} \int {d^{4} \ell \over (2\pi)^{4}}
~\delta^{(+)}(\ell^2)~\delta^{(+)}((\ell-K)^2)~{\cal
T}^{(N)}(\ell) , \eea
where the integrand can be generally represented as
\bea {\cal T}^{(N)}(\ell) = \frac{ \prod_{j=1}^{n+k}
(-2\ell\cdot P_j)}{\prod_{i=1}^k D_i(\ell) },~~~~ D_i(\ell)=(\ell-K_i)^2 \eea
Here $N$ is defined as the degree of amplitude, which is just equal to $n$ and is the half power of momentum in the fraction. To carry out the integration, we use the spinor technique to write the loop momentum as
 $\ell=t\la\tl$, with
$\lambda, \tl \in \IC \IP^1 $, and  the measure can express as following
\bea
\int d\mu ~~ (\bullet )& = & \int d^4 \ell \delta(\ell^2) \delta((\ell-K)^2) ~~ (\bullet )\nonumber \\
& = &  \int_0^\infty t dt \int
\braket{\la~d\la}[\W \la~d\W \la]  \delta( K^2- t \gb{\la|K|\tl})~~ (\bullet ) \nonumber  \\
& = & \int_{\bbar\la=\tl}
\braket{\la~d\la}[\W \la~d\W \la]
{K^2\over \gb{\la|K|\tl}^2} ~~ (\bullet )
\label{lips}
\eea
where the $t$ integration has been carried out. After the $t$-integration, the ${\cal T}^{(N)}(\ell)$ becomes
\bea {\cal T}^{(N)}(\la,\tl )  = {(K^2)^{n}\over
\gb{\la|K|\tl}^{n}} {\prod_{j=1}^{n+k} \gb{\la|R_j|\tl}\over
\prod_{i=1}^k \gb{\la|Q_i|\tl}}, \quad  \eea
where $Q_i=\frac{K_i^2}{K^2}K-K_i$ and $R_i=-P_i$. We also define the integrand
\bea
I_{\rm term}={K^2\over \gb{\la|K|\tl}^2}  {\cal T}^{(N)}(\la,\tl )= \frac{G(\la)\prod_{j=1}^{n+k}[a_j|\tl]}{\gb{\lambda|K|\tl}^{n+2}\prod_{i=1}^k\gb{\lambda|Q_i|\tl}},
\label{iterm}
\eea
where $G(\la)$ is constant, and $[a_j|=\bra{\la|R_j}$.

The expression ${\cal T}^{(N)}(\la,\tl )$ contains all information of coefficients of  boxes, triangles and
bubbles. To disentangle the information, canonical splitting has been given in \cite{Britto:2006fc,Britto:2007tt,Britto:2008sw,Britto:2008vq,Feng:2008ju}.
\begin{equation}
    \frac{[a| \tl]}{\gb{\lambda|Q_1|\tl}\gb{\lambda|Q_2|\tl}}=\frac{\tgb{a|Q_1|\lambda}}{\braket{\lambda|Q_2Q_1|\lambda}\gb{\lambda|Q_1|\tl}}+\frac{\tgb{a|Q_2|\lambda}}{\braket{\lambda|Q_1Q_2|\lambda}\gb{\lambda|Q_2|\tl}}
\end{equation}
After making the splitting, we get the canonical splitting
\cite{Britto:2007tt}
\bea
    I_{\rm term}& =&\sum_{i=1}^{n+1}\lim_{s_i\to0 }\frac{1}
    {\gb{\lambda| K|\tl}\gb{\lambda|K+s_i\eta|\tl}}\frac{G(\la)
    \prod_{j=1}^{n+k}\gb{a_j|K+s_i\eta|\tl}}{\prod_{q\neq i}^{n+1}
    \gb{\lambda|(K+s_q\eta)(K+s_i\eta)|\tl}\prod_{p=1}^k\gb{\lambda|Q_p(K+s_i\eta)|\tl}}
     \nonumber \\
     & +&\sum_{i=1}^{k}\frac{1}{\gb{\lambda|K|\tl}\gb{\lambda|Q_i|\tl}}
      \frac{G(\la)\prod_{j=1}^{n+k} \tgb{a_j|Q_i|\lambda}}
      {\braket{\lambda |K Q_i|\lambda}^{n+1}\prod_{r\neq i}^{r=k}
      \braket{\lambda |Q_r Q_i|\lambda}}.~~~\label{cano-split}
\eea
The second line contains all information of coefficients of triangle
and boxes, while the first line contains purely the information of
coefficients of bubbles. More explicitly, by taking residues of
various poles at the first line, we get the bubble coefficients.
Based on above canonical splitting,  we can extract coefficients of
various master integrals. The algebraic expressions are summarized
as following\cite{Britto:2010xq}:

\begin{itemize}

\item {\bf Box coefficients}

The coefficient of the box, identified by the two cut propagators along with $D_r$ and $D_s$, is given by
\bea
\hspace{-0.5in}
  C[K_r,K_s,K]  =  {1\over 2}
\left.\left(
{\cal T}^{(N)}(\ell) D_r(\ell) D_s(\ell)
\right|_{\la \to P_{sr,1},\tl \to P_{sr,2}}
+ \{P_{sr,1}\leftrightarrow P_{sr,2}\}\right)
\Label{box-formula}
\eea
We define $Q_s$ and $Q_r$ as following
\begin{equation}
    \begin{split}
        Q_s=\frac{K_s^2}{K^2}K-K_s\\
        Q_r=\frac{K_r^2}{K^2}K-K_r
    \end{split}
\end{equation}
And define auxiliary vectors $P_{sr,1}$ and $P_{sr,2}$ are the null linear combinations of $Q_r$ and $Q_s$.
\bea
P_{sr,1} &=& Q_s + \left( {-Q_s \cdot Q_r + \sqrt{\Delta_{sr}}\over Q_r^2} \right) Q_r,\nonumber \\
P_{sr,2} &=& Q_s + \left( {-Q_s \cdot Q_r - \sqrt{\Delta_{sr}}\over Q_r^2} \right) Q_r,\nonumber \\
\Delta_{sr} &=& (Q_s \cdot Q_r)^2- Q_s^2 Q_r^2.
\label{eq:def4}
\eea

\item {\bf Triangle coefficients}

If $N <-1$, the triangle coefficients are zero.  If $N \geq - 1$, the coefficient of the triangle, identified by the two cut propagators along with  $D_s$, is given by
\bea
\label{eqn:tri}
\hspace{-0.5in}
C [K_s,K]  &=&
  {1\over 2(N+1)!\sqrt{\Delta_s}^{N+1}
\vev{P_{s,1}~P_{s,2}}^{N+1}}
\\ & & \times \frac{d^{N+1}}{d\tau^{N+1}}
\left.\left(\left.
{\cal T}^{(N)}(\ell) D_s(\ell) \gb{\la|K|\tl}^{N+1}
\right|_{\tl \to Q_s \la ,\la \to P_{s,1}- \tau P_{s,2}}
\right.\right.
\nonumber \\ & &
~~~~~~~~~~~~~~~~+ \{P_{s,1}\leftrightarrow P_{s,2}\}
\Bigg)
\Bigg|_{\tau \to 0}
\nonumber
\eea
Here we use the following definitions. The vectors $P_{s,1}$ and $P_{s,2}$ are null linear combinations of $Q_s$ and $K$.
\bea
P_{s,1} &=& Q_s + \left({-Q_s \cdot K + \sqrt{\Delta_{s}}\over K^2} \right) K,\nonumber
 \\
P_{s,2} &=& Q_s + \left({-Q_s \cdot K - \sqrt{\Delta_{s}}\over K^2} \right) K,\nonumber
\\ \Delta_{s} &=& (Q_s \cdot K)^2- Q_s^2 K^2.
\label{eq:def3}
\eea
The effect of the multiple derivative of the parameter $\tau$, evaluated at $\tau=0$, is simply to pick out a term in the series expansion.

\item {\bf Bubble coefficients}

There is just one bubble in the cut channel $K$.  If $N<0$, the coefficient is zero.  If $N \geq 0$, the coefficient is
\bea
C[K] = K^2  \sum_{q=0}^N {(-1)^q\over q!} {d^q \over
ds^q}\left.\left( {\cal B}_{N,N-q}^{(0)}(s)
  + \sum_{r=1}^k\sum_{a=q}^N
\left({\cal B}_{N,N-a}^{(r;a-q;1)}(s)-{\cal
B}_{N,N-a}^{(r;a-q;2)}(s)\right)\right)\right|_{s=0},
\eea
where
\bean
 {\cal B}_{N,m}^{(0)}(s)\equiv {d^N\over d\tau^N}\left.\left(
\left.
{(2\eta\cdot
K)^{m+1}  \gb {\la|K|\tl}^N  \over N! [\eta|\eta' K|\eta]^{N}(m+1) (K^2)^{m+1} \vev{\la~\eta}^{N+1} }
{\cal T}^{(N)}(\ell)
\right|_{\stackrel{\tl \to (K+s \eta)\cdot\la}{\la \to (K-\tau \eta')\cdot\eta}}
\right)
\right|_{\tau \to 0},
\eean
\bean
 {\cal B}_{n,m}^{(r;b;1)}(s)  \equiv  & & {(-1)^{b+1}\over
 b! (m+1) \sqrt{\Delta_r}^{b+1} \vev{P_{r,1}~P_{r,2}}^b} \times
\\ & &
{d^b \over d\tau^{b}}
\left.
\left( {\gb{\la|\eta|P_{r,1}}^{m+1}
\vev{\la|Q_r \eta|\la}^{b} \gb {\la|K|\tl}^{N+1}
\over
\gb{\la|K|P_{r,1}}^{m+1} \vev{\la|\eta K|\la}^{n+1} }
{\cal T}^{(N)}(\ell) D_r(\ell)
\right)
\right|_{\stackrel{\tl \to (K+s \eta)\la,~\la \to P_{r,1}-\tau P_{r,2}}{\tau=0}}
\eean
\bean
 {\cal B}_{n,m}^{(r;b;2)}(s)  \equiv  & & {(-1)^{b+1}\over
 b! (m+1) \sqrt{\Delta_r}^{b+1} \vev{P_{r,1}~P_{r,2}}^b} \times
\\ & &
{d^b \over d\tau^{b}}
\left.
\left( {\gb{\la|\eta|P_{r,2}}^{m+1}
\vev{\la|Q_r \eta|\la}^{b} \gb {\la|K|\tl}^{N+1}
\over
\gb{\la|K|P_{r,2}}^{m+1} \vev{\la|\eta K|\la}^{n+1} }
{\cal T}^{(N)}(\ell) D_r(\ell)
\right)
\right|_{\stackrel{\tl \to (K+s \eta)\la,~\la \to P_{r,2}-\tau P_{r,1}}{\tau=0}}
\eean
Here $\eta,\eta'$ are arbitrary spinors which should be generic in the sense that they do not coincide with any spinors from massless external legs.

Above expressions for bubble coefficients look complicated.
However, it is just the calculations of residues of various
poles in the first line of canonical splitting
(\ref{cano-split}). As it will be clear in the section three,
our comparison will be done at the level of (\ref{cano-split})
only.

\end{itemize}

\subsection{Review of generalized unitarity cut method}
\label{sec:gm}

As we have mentioned, our purpose in this paper is to  establish the
explicit relation between unitarity cut method and the generalized
unitarity cut method proposed in\cite{ArkaniHamed:2008gz} . In this
subsection, we will briefly review their results.

The key idea of their method is the generalization of formula (\ref{eq:mi}) with multiple cut
\bea
\text{Cut}^{(n)}_{\{1\},\dots,\{n\}} A^{\text{1-loop}}=\sum_c C_d^c\text{Cut}^{(n)}_{\{1\},\dots,\{n\}}I_d^c+\cdots+\sum_c C_2^c\text{Cut}^{(n)}_{\{1\},\dots,\{n\}}I_2^c,
\eea
where the sum of $c$ is over all different channels, and $\text{Cut}^{(n)}_{\{1\},\dots,\{n\}}$ means to cut $n$ propagators $D_1,\dots,D_n$.
Based on this formula, when applying to $D=4$,
coefficients of master integrals can be read out as following:
\begin{itemize}

\item {\bf Box coefficients}

For quadruple cut, $\text{Cut}^{(4)}$ is simply proportional to $C_4$. That's because quadruple cut renders $I_n(n<4)$ zero\cite{Britto:2004nc,Bern:2004ky} . To be concrete,
\bea
C_4 =
\frac{1}{2} \sum_{\ell \in \cal S} A_{L,s,1}(\ell)A_{L,s,2}(\ell) A_{R,r,1}(\ell)A_{R,r,2}(\ell),
\label{qcutboxcoeff}
\eea
where $\cal S$ is the solution set for the four delta functions of the cut propagators
\bea
\hspace{-0.5in}
{\cal S}=\{\ell~ | \ell^2=0, \, (\ell-K)^2=0, \, (\ell
-K_s)^2=0, \, (\ell-K_r)^2=0\}.
\label{qcut}
\eea

\item {\bf Triangle coefficients}

For triangle coefficients in 4D, it's not so lucky
because triple cuts can not fix the internal momentum completely and there is a free parameter left. A consequence of this freedom is that  some box integrals will contribute to triple cuts.
Thus we need to have a cleverer way to disentangle their information. The way to do so is following \cite{ArkaniHamed:2008gz}.   Suppose we cut propagators $\ell_3^2=\ell^2,\ell_2^2=(\ell-K_s)^2,\ell_1^2=(\ell-K)^2$. Without loss of generality, we can choose external condition with a Lorentz boost to be
\bea
K_{a\dot{a}}&=&\left( \begin{array}{cc} -1 & 0
    \\ 0 & -1 \end{array} \right), \nonumber \\
(K_s)_{a\dot{a}}&=&\left(\begin{array}{cc} E_{+} & 0
    \\ 0 & E_{-} \end{array} \right),
\label{ec}
\eea
where $P_{a\dot{a}}=p_\mu (\bar \sigma^\mu)_{a\dot{a}}=\left( \begin{array}{cc} p^0+p^3& p^1-ip^2
\\p^1+ip^2 & p^0-p^3\end{array} \right)$.
Now the on shell condition gives \bea \ell_i=
\left(\begin{array}{cc} \alpha_i^{+} & \bar l \\ l &
\alpha_i^{-} \end{array} \right) =\left(\begin{array}{cc}
\alpha_i^{+} & re^{-i\theta} \\ re^{i\theta} & \alpha_i^{-}
\end{array} \right)~~~\label{Nima-ell} \eea with \bea l\bar
l=\alpha_i^+ \alpha_i^-\equiv
r^2=\frac{-E_+E_-(1+E_+)(1+E_-)}{(E_+-E_-)^2}, \eea and
$\alpha^\pm_i$ are some definite functions of $E_\pm$ whose
explicit expressions are not important in the derivation. Here
we only discuss under condition $r^2>0$, while the result of
other regions of $r^2$ can be obtain by analytic continuation.
Integrate out all delta functions, we have \bea \text{Cut}^{(3)}
\propto \int_0^{2\pi} d \theta F(r\cos\theta,r\sin\theta), \eea
where $F=A_1A_2A_3$ is the factorized tree amplitude after
triple cut. To find the proportionality constant, we consider
the simplest case when loop amplitude \bea A^{\rm one-loop}=\int
\frac{d^{4}\ell}{(2\pi)^{4}}
\frac{1}{\ell^2(\ell+P_1)^2(\ell-P3)^2}. \eea We expect
$\text{Cut}^{(3)}=C_3=1$, because no box integrals exist. Now
$A_1A_2A_3=1$ and we find the proportionality constant equals to
$1/2\pi$.

For further derivation, we change the variable to $z=\cos{\theta}$. The integral becomes
\bea
\text{Cut}^{(3)} =\frac{1}{2\pi} \int_{-r}^{r}\frac{d z}{\sqrt{r^2 - z^2}}\left[F(z,\sqrt{r^2 - z^2}) + F(z,-\sqrt{r^2 - z^2})\right].
\eea
Then we consider $z$ to be a complex variable. That's because we want to identify the box integral contribution from $\text{Cut}^{(3)}$, and the momenta satisfy quadruple cut are always complex. The integrand has a branch cut which can be taken to be $(-r,r)$, and the integral itself can be rewritten to be a contour integral encircling the branch cut clockwisely
\bea
\text{Cut}^{(3)} = \frac{1}{4\pi}\int_{{\cal C}_0} \frac{d z}{\sqrt{r^2 - z^2}} \left[F(z,\sqrt{r^2 - z^2}) + F(z,-\sqrt{r^2 - z^2})\right].
\label{eq:cut3}
\eea
The integrand of above integral has simple poles on complex $z$ plane. Simple poles at finite $z$ comes from the remaining propagators in $A_1(z)A_2(z)A_3(z)$. When $z$ approaches these poles, an additional propagator goes on shell, and the integrand further factorized to be a product of four pieces of tree amplitudes $A'_1(z) A'_2(z) A'_3(z) A'_4(z)$, which is proportional to $C_4^c$ for some channel $c$. It's now clear that to eliminate the contributions from box integrals, we can simply drop the residues at finite poles, and take the residue at infinity only. So we deform the contour to obtain
\bea
C_3 = \frac{1}{4\pi}\int_{{\cal C}} \frac{d z}{\sqrt{r^2 - z^2}} \left[F(z,\sqrt{r^2 - z^2}) + F(z,-\sqrt{r^2 - z^2})\right],
\label{eq:c3}
\eea
where the contour ${\cal C}$ encircles the pole at infinity.

\item {\bf Bubble coefficients}

For bubble coefficients, as triple cut case, double cut will
contain contributions from boxes and triangles. The  separation  of bubble part from others is done in \cite{ArkaniHamed:2008gz} as following

\begin{equation}
C_2 = \int d {\rm LIPS}[\ell_1,\ell_2] \int_{{\cal C}} \frac{dz}{z}
M_L(\ell_1(z),\ell_2(z)) M_R(\ell_1(z),\ell_2(z)),\label{C2}
\end{equation}
where we deformed the loop momentum using the BCFW shift \cite{Britto:2004ap,Britto:2005fq} with $\ell_1=\ell_1-z q$ and $\ell_2=\ell_2+z q$ with $q$ a reference momentum, which keeps $\ell_1,\ell_2$ on shell as well as momentum conservation. The contour $\cal C$ encircles the pole at infinity.

\end{itemize}
%

\section{Connection between these two methods}

Having roughly reviewed two methods in previous section, in this
section we will show their connection explicitly.

\subsection{Connection between input amplitude}

The main difference of these two methods is the number of cut
implement  on the loop propagators. For unitarity cut method with
double cuts, the input is always multiplication of two on-shell tree
level amplitudes, i.e.,
\bea
{\rm Input_2} ={\cal T}^{({N})}(\ell) ={A_L A_R}
\eea
where we have assumed the propagators $\ell^2,(\ell-K)^2$ have been
cut. However, for generalized unitarity cut method, depending on
which coefficient we are looking for, the input is different.  For
triangle coefficient, the triple cut is needed and the input is
\bea
{\rm Input_3} ={A_{L,s,1}A_{L,s,2} A_R},
\eea
where an extra propagator $D_s=(\ell-K_s)^2$ has been cut. For box coefficient, the quadruple cut is needed and the input is
\bea
{\rm Input_4} ={A_{L,s,1}A_{L,s,2} A_{R,r,1}A_{R,r,2}}
\eea
where two extra propagators $D_s=(\ell-K_s)^2$ and $D_r=(\ell-K_r)^2$ have been cut.

Though the input amplitudes seem to be different, they can be easily related by the factorization property of tree amplitude. For example,
\bea
A_L \times A_R\times D_s \Bigg|_{
    D_s=0} =\sum_c {A_{L,c,1}A_{L,c,2} A_R\over D_c} \times D_s\Bigg|_{
    D_s=0} ={A_{L,s,1}A_{L,s,2} A_R}
\eea

After calculating each term in the formula, it's easily to show the relationship of the left and right side of the equation.
\bea
{\cal T}^{({N})}(\ell)&=& A_L\times A_R
\nonumber\\
{\cal T}^{({N})}(\ell)\cdot D_{s}(\ell)\Bigg|_{D_s=0}&=& A_{L,s,1}A_{L,s,2} A_R
\nonumber\\
\label{eq:input}
{\cal T}^{({ N})}(\ell) \cdot D_s(\ell) \cdot D_r(\ell) \Bigg|_{D_s,D_r=0} &=&  A_{L,s,1}A_{L,s,2} A_{R,r,1}A_{R,r,2}
\eea
These relations will be used when we prove the equivalence of two methods.

\subsection{The equivalence of box coefficient between two methods}

Let us recall the box coefficients of unitarity cut method in (\ref{box-formula})
\bea C[K_r,K_s,K] & = & {1\over 2}\left({\cal
T}^{({N})}(\ell) \cdot D_r(\ell) \cdot D_s(\ell) \Bigg|{\begin{matrix} \\
\left\{\scriptsize \begin{matrix} \bket{\tl} & \rightarrow &
\bket{P_{rs,2}}
\\ \ket{\la} & \rightarrow  & \ket{P_{rs,1}}
\end{matrix}\right. \end{matrix}} +
\{P_{rs,1}\leftrightarrow P_{rs,2}\} \right)~,~~~\label{box-1}\qquad
\eea
The role of replacement of $\bket{\tl}$ and $\ket{\la}$ is to put propagators $D_r(\ell)$ and $D_s(\ell)$ on shell. To see it, one can see that using the definition in (\ref{eq:def4}), $Q_s$ and $Q_r$ can be expressed in terms of $P_{rs,1}$ and $P_{rs,2}$ as
\bea
Q_{s} &=&{ {\sqrt{\Delta_{rs}}+Q_s\cdot Q_j}\over{2 \sqrt{\Delta_{rs}}} }P_{rs,1}+{{\sqrt{\Delta_{rs}}-Q_s\cdot Q_j}\over{2 \sqrt{\Delta_{rs}}}}P_{rs,2}
\nonumber \\
Q_{r} &=& {Q_r^2\over{2\sqrt{\Delta_{rs}}}}(P_{rs,1}-P_{rs,2})
~~~~\label{Qij}
\eea
Thus we have
\bea
\gb{\la|Q_s|\tl}&=&{ {\sqrt{\Delta_{rs}}+Q_s\cdot Q_r}\over{2 \sqrt{\Delta_{rs}}} }\gb{\la|P_{rs,1}|\tl}+
{ {\sqrt{\Delta_{rs}}-Q_s\cdot Q_r}\over{2 \sqrt{\Delta_{rs}}} }\gb{\la|P_{rs,2}|\tl}
\nonumber\\
\gb{\la|Q_r|\tl}&=& {Q_r^2\over{2\sqrt{\Delta_{rs}}}}(\gb{\la|P_{rs,1}|\tl}-\gb{\la|P_{rs,2}|\tl})
~~~~\label{Qij}
\eea
which are zero after the substitutions $\bket{\tl}\rightarrow\bket{P_{rs,2}},\ket{\la}\rightarrow\ket{P_{rs,1}}$ or $\bket{\tl}\rightarrow\bket{P_{rs,1}},\ket{\la}\rightarrow\ket{P_{rs,2}}$. Namely, propagators
\bea
D_s&=&(\ell-K_s)^2 = { K^2 \gb{\la|Q_s|\tl} \over \gb{\la|K|\tl} }=0
\nonumber\\
D_r&=&(\ell-K_r)^2 = { K^2 \gb{\la|Q_r|\tl} \over \gb{\la|K|\tl} }=0
\eea
are indeed on shell after above substitutions. With this observation and (\ref{eq:input}), the box coefficient (\ref{box-1}) is simply
\bea
C[K_r,K_s,K] =
 \frac{1}{2} \sum_{\ell \in \cal S} A_{L,s,1}(\ell)A_{L,s,2}(\ell) A_{R,r,1}(\ell)A_{R,r,2}(\ell),
\label{qcutboxcoeff-1} \eea
which is the same as result (\ref{qcutboxcoeff}) coming from
generalized unitarity cut method. Thus we have proved the expression
of box coefficient for unitarity cut method and generalized
unitarity cut method are indeed same.

\subsection{Triangle coefficients}

For triangle coefficient, the connection between the two methods is
not  so obvious. The main difference is the choice of parameter for
cut on-shell complexified momentum. In unitarity cut method, we
parameterize the momentum as $\ket{\ell} \rightarrow
\ket{P_{s,1}-\tau P_{s,2}}$, where a complex parameter $\tau$ is
introuduced. In generalized unitarity cut method,  after putting
three propagators on shell, we introduce a degree-of-freedom
parameter $z$. Though $\tau$ and $z$ seem to be very different,
there is a simple relation between these two parameters, which we
will derive by direct calculation. Using  this relation, we will
prove the equivalence of unitarity cut method and generalized
unitarity cut method on computing triangle coefficient.

\subsubsection{Connection between parameter $\tau$ and $z$}

To find the connection between $\tau$ and $z$, we notice that they both show up in loop momentum $\ell$. We first calculate $\ell$ by unitarity cut method.  For comparison with \cite{ArkaniHamed:2008gz}, we put $\ell^2=(\ell-K)^2=(\ell-K_s)^2=0$ and use the external condition (\ref{ec}), thus
 by the definition (\ref{eq:def3}) of $Q_s$, $P_{s,1}$ and $P_{s,2}$, we obtain
\bea
Q_s&=&\frac{K_s^2}{K^2}K-K_s\\\nonumber
&=&\left( \begin{array}{cc} -E_{+}E_{-}-E_{+} & 0
    \\0& -E_{+}E_{-}-E_{-} \end{array} \right)\equiv \left(\begin{array}{cc} \alpha_s^{+} & 0 \\ 0 & \alpha_s^{-} \end{array} \right)\\\nonumber
P_{s,1}&=& \left(\begin{array}{cc} \alpha_s^{+}-\alpha_s^{-} & 0 \\ 0 & 0 \end{array} \right)=\left(\begin{array}{cc} 1 \\ 0 \end{array} \right)\left(\begin{array}{cc} \alpha_s^{+}-\alpha_s^{-} & 0 \end{array} \right)\\\nonumber
P_{s, 2}&=& \left(\begin{array}{cc} 0 & 0 \\ 0 & \alpha_s^{-}-\alpha_s^{+} \end{array} \right)=\left(\begin{array}{cc} 0 \\ -1 \end{array} \right)\left(\begin{array}{cc} 0 & \alpha_s^{+}-\alpha_s^{-} \end{array} \right)
\eea
Since we have $\ell_{a\dot{a}}=t\la_a\W\la_{\dot a}$ and the substitution $\bket{\W\la}=Q_s\ket{\la}$ and $\ket{\la}=\ket{P_{s,1}-\tau P_{s,2}}$, we can derive
\bea
\ket{\la}_a= \ket{P_{s,1}}_a - \tau\ket{P_{s,2}}_a&=& \left(\begin{array}{cc} 1 \\ \tau \end{array} \right)\\\nonumber
{\left[ \W\la\right|}_{\dot a}= \eps_{\dot a \dot b}(Q_s)^{\dot b b}\ket{\la}_b&=&\left(\begin{array}{cc} \tau \alpha_s^{+} & -\alpha_s^{-} \end{array}\right) \\\nonumber
t ={K^{2} \over \gb{\la|K|\la}}&=&\frac{1}{\tau(\alpha_s^{-}-\alpha_s^{+})}
\eea
and finally
\bea
\ell&=&t\ket{\la}_a{\left[ \W\la\right|}_{\dot a}= \frac{1}{\alpha_s^{-}-\alpha_s^{+}}\left(\begin{array}{cc} \alpha_s^{+} & \frac{-\alpha_s^{-}}{\tau} \\ \alpha_s^{+}\tau & -\alpha_s^{-} \end{array} \right) \nonumber\\
&=& \left(\begin{array}{cc} \alpha_3^{+} & \bar l \\ l & \alpha_3^{-} \end{array} \right)
\eea
where $l={\alpha_s^{+} \over {\alpha_s^{-}-\alpha_s^{+}}}\tau$ and $\bar l={-\alpha_s^{-} \over {\alpha_s^{-}-\alpha_s^{+}}}\frac{1}{\tau}$. Notice that $\alpha_3^{+}$ and $\alpha_3^{-}$ define above is the same as formula (140) in \cite{ArkaniHamed:2008gz}, which shows the on-shell condictions are satisfied. Similarly, for the substitution $\bket{\W\la}=Q_s\ket{\la}$ and $\ket{\la}=\ket{P_{s,2}-\tau P_{s,1}}$, we have $l={\alpha_s^{+} \over {\alpha_s^{-}-\alpha_s^{+}}}\frac{1}{\tau}$ and $\bar l={-\alpha_s^{-} \over {\alpha_s^{-}-\alpha_s^{+}}}\tau$.

Next we consider generalized unitarity cut method. Comparing to the
definition of loop momentum $\ell$ in (\ref{Nima-ell}), we find the
relation between parameter $\tau$ and $z$ \bea z =r\cos{\theta}
=\frac{\tilde{l}+l}{2}=\frac{1}{2(\alpha_s^--\alpha_s^+)}(\alpha_s^+\tau+\frac{-\alpha_s^-}{\tau})
\label{eq:z_to_tau} \eea for the substitution
$\ket{\la}=\ket{P_{s,1}-\tau P_{s,2}}$. For another substitution
$\ket{\la}=\ket{P_{s,2}-\tau P_{s,1}}$, it is just to put
$\tau\rightarrow\frac{1}{\tau}$.

\begin{figure}
    \centering
    \includegraphics[width=18cm, height=6cm]{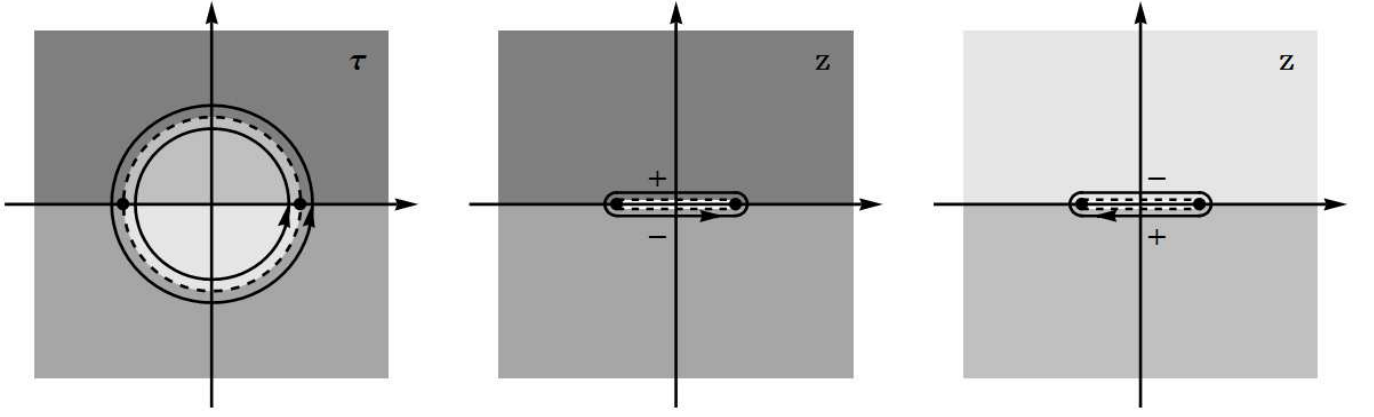}
    \caption{Joukowsky transformation}\label{fig:tau2z}
\end{figure}

It's useful to figure out how the complex plane transform when we change the variable by formula (\ref{eq:z_to_tau}). Remember that following \cite{ArkaniHamed:2008gz}, we suppose $r^2=-\alpha_s^+\alpha_s^-/(\alpha_s^--\alpha_s^+)^2>0$ all along our derivation, while the result of $r^2\le 0$ can be obtained by analytic continuation. By rescaling $\tau$ and $z$, (\ref{eq:z_to_tau}) is identical to the well-known Joukowsky transformation
\bea
z =\frac{1}{2}(\tau+\frac{1}{\tau}).
\label{eq:jtrans}
\eea
This transformation maps the whole $\tau$ plane to two sheets of $z$ plane (because $\tau=z\pm\sqrt{z^2-1}$), $\abs{\tau}>1$ to one and $\abs{\tau}<1$ to another, as figure \ref{fig:tau2z} shows. The points on $z$ plane are branch points, and the dashed lines are branch cuts. We choose $\tau=z+\sqrt{z^2-1}$, as well as $Arg(z)=0$ on the $x$ axis of the first sheet. Under this convention, the $+$ and $-$ in the figure means $\sqrt{1-z^2}$ to be positive or negative near the branch cut.

Notice that function $f(z,\sqrt{1-z^2})$ and $f(z,-\sqrt{1-z^2})$ can be viewed as a same function $f(z)$ on different sheets. Thus we can rewrite formulas (\ref{eq:cut3}) to be
\bea
\text{Cut}^{(3)} = \frac{1}{4\pi}\int_{{\cal C}} \frac{d z}{-\sqrt{1 - z^2}} A_1 A_2 A_3(z) = \frac{1}{4\pi}\int_{{\cal C'}} \frac{d \tau}{i \tau} A_1 A_2 A_3(\tau)
\label{eq:newcut3}
\eea
where the contour $\cal C$ on $z$ plane and the corresponding contour $\cal C'$ on $\tau$ plane are shown on figure \ref{fig:tau2z}. To rescale back we simply put $\sqrt{1 - z^2}\rightarrow\sqrt{r^2 - z^2}$ in the integrand.

\subsubsection{Connection between residues}
Now we come to the explicit formula of triangle coefficient. In unitarity cut method, the triangle coefficient comes from
\bea
\label{eqn:tri}
\hspace{-0.5in}
C [K_s,K]  &=&
{1\over 2(N+1)!\sqrt{\Delta_s}^{N+1}
    \vev{P_{s,1}~P_{s,2}}^{N+1}}
\\ & & \times \frac{d^{N+1}}{d\tau^{N+1}}
\left.\left(\left.
{\cal T}^{(N)}(\ell) D_s(\ell) \gb{\la|K|\tl}^{N+1}
\right|_{\tl \to Q_s \la ,\la \to P_{s,1}- \tau P_{s,2}}
\right.\right.
\nonumber \\ & &
~~~~~~~~~~~~~~~~+ \{P_{s,1}\leftrightarrow P_{s,2}\}
\Bigg)
\Bigg|_{\tau \to 0}
\nonumber
\eea

It can be easily shown
\bea
\sqrt{\Delta_s}&=&\alpha_s^{-}-\alpha_s^{+},\\\nonumber
\vev{P_{s,1}~P_{s,2}}&=&1,\\\nonumber
\gb{\la|K|\tl}&=&\tau(\alpha_s^{-}-\alpha_s^{+}).
\eea
Note that these equations are valid for two substitutions.

We concentrate on the first term in the derivative. Under the
substitution $\bket{\la}=Q_s\ket{\la}$ and
$\ket{\la}=\ket{P_{s,1}-\tau P_{s,2}}$, we put the propagator
$1/D_s(\ell)$ on shell. Using the fact that ${\cal T}^{(N)}(\ell)
D_s(\ell)$ has a Laurent expansion $\sum_{n=-(N+1)}^\infty {a_n
\tau^n}$ around $\tau=0$, we can conclude that \bea
&&\left.\frac{1}{(N+1)!\sqrt{\Delta_s}^{N+1}\vev{P_{s,1}~P_{s,2}}^{N+1}}
\frac{d^{N+1}}{d\tau^{N+1}}{\cal T}^{(N)}(\ell) D_s(\ell)
\gb{\la|K|\tl}^{N+1} \right|_{\tl \to Q_s \la ,\la \to P_{s,1}- \tau
P_{s,2}}\\\nonumber &=&\left.
\frac{1}{(N+1)!}\frac{d^{N+1}}{d\tau^{N+1}}\sum_{n=-(N+1)}^\infty
{a_n \tau^n}\cdot\tau^{N+1} \right|_{\tau\to 0} = a_0 \\\nonumber
&=&\text{Res}_{\tau=0}(\frac{{\cal T}^{(N)}(\tau)\times
D_s(\tau)}{\tau})\\\nonumber &=&\text{Res}_{\tau=0}(\frac{A_1 A_2
A_3}{\tau}), \eea where at the last equal sign we use
(\ref{eq:input}), and we denote the factorized tree amplitude
$A_{L,s,1}A_{L,s,2} A_R$ by $A_1 A_2 A_3$. For the second term in
the derivative, we simply set $\tau\rightarrow 1/\tau$ in ${\cal
T}^{(N)}(\ell) D_s(\ell)$, while $\gb{\la|K|\tl}$ remains
proportional to $\tau$. We obtain \bea
&&\left.\frac{1}{(N+1)!\sqrt{\Delta_s}^{N+1}\vev{P_{s,1}~P_{s,2}}^{N+1}}
\frac{d^{N+1}}{d\tau^{N+1}}{\cal T}^{(N)}(\ell) D_s(\ell)
\gb{\la|K|\tl}^{N+1} \right|_{\tl \to Q_s \la ,\la \to P_{s,2}- \tau
P_{s,1}}\\\nonumber &=&\text{Res}_{\tau=0}(\frac{{\cal
T}^{(N)}(\frac{1}{\tau})\times
D_s(\frac{1}{\tau})}{\tau})=-\text{Res}_{\tau=\infty}(\frac{{\cal
T}^{(N)}(\tau)\times D_s(\tau)}{\tau})\\\nonumber
&=&-\text{Res}_{\tau=\infty}(\frac{A_1 A_2 A_3}{\tau}), \eea where
we use $\text{Res}_{z=\infty}
f(z)=\text{Res}_{z=0}(-\frac{f(1/z)}{z^2})$.

Combining above results,
\bea
\label{eq:umres}
C [K_s,K] = \frac{1}{2}\text{Res}_{\tau=0}(\frac{A_1 A_2 A_3}{\tau})-\frac{1}{2}\text{Res}_{\tau=\infty}(\frac{A_1 A_2 A_3}{\tau}).
\eea
It's now clear that what formula (\ref{eqn:tri}) really do is to compute the residue of $\frac{A_1 A_2 A_3}{\tau}$ at $\tau=0$ and $\tau=\infty$.

After figuring out the meaning of formula (\ref{eqn:tri}), we turn
to the generalized unitarity cut method. We have already shown that
triple cut can be computed by (\ref{eq:newcut3}). \bea
\label{eq:gumres} \text{Cut}^{(3)} = \frac{1}{4\pi}\int_{{\cal C}}
\frac{d z}{-\sqrt{r^2 - z^2}} A_1 A_2 A_3(z) = \frac{1}{4\pi
i}\int_{{\cal C'}} \frac{d \tau}{\tau} A_1 A_2 A_3(\tau) \eea To get
rid of the influence of box coefficient, we stretch the contour
$\cal C$ to be two infinitely large loops, which only contain the
residue on $z=\infty$. Now, the contour $\cal C'$ becomes an
infinitesimal loop encircling $\tau=0$ and a infinitely large loop
encircling $\tau=\infty$. Thus we come to the final step \bea C_3=i
\text{Res}_{z=\infty\text{ on two sheets}}(\frac{A_1 A_2
A_3}{-\sqrt{r^2 - z^2}})=\frac{1}{2}\text{Res}_{\tau=0}(\frac{A_1
A_2 A_3}{\tau})-\frac{1}{2}\text{Res}_{\tau=\infty}(\frac{A_1 A_2
A_3}{\tau}) \eea

We have proven the formulas for triangle coefficient in two methods are indeed equal.

\subsection{Bubble coefficients}
%
The core part in evaluating the bubble coefficient of master
integral is to split it from other master integrals. In unitarity
cut method, the procedure is done by recognizing it from analytical
property of other  master integrals, which behave as  pure
logarithm. In generalized unitarity cut method,  it is done by
recognizing it as the infinite pole in the
integrand\cite{ArkaniHamed:2008gz}.

We begin with expression (\ref{C2})
\begin{equation}
    C_2=\int {d\rm LISP} \int_{\cal{C}} \frac{dz}{z } A_L(z)\times A_R(z),~~~\label{gen-bubble-ceoff}
\end{equation}
However,  after two cuts, the contribution coming from triangles and boxes will appear as some remaining propagators in the form of $\frac{1}{(l-K_i)^2}$. When we do the contour integrals at infinity, i.e., when the contour $\cal C$ is a infinitely large loop, we have
\begin{equation}
    \int_{\cal C} \frac{dz}{z}\mathcal{T}^{(N)}(z)=2\pi i\mathcal{T}^{(N)}(0)+2\pi i \sum_i ~~\frac{1}{z_i}{\rm Res}_{z=z_i}\mathcal{T}^{(N)}(z) ,
\end{equation}
with some $z_i$ that put a remaining propagator $D_i$ on shell. Rewriting above formula as
\bea \mathcal{T}^{(N)}(0)=-\sum_i ~~\frac{1}{z_i}{\rm Res}_{z=z_i}\mathcal{T}^{(N)}(z)+{1\over 2\pi i}\int_{\cal C} \frac{dz}{z}\mathcal{T}^{(N)}(z). \eea
Since we can exchange the order of integration, we integrate $t$ in $d\rm LISP$ as in (\ref{lips}), then the integrand becomes (\ref{iterm}). Thus we have
\begin{equation}
    I_{\rm term}(0)=-\sum_i ~~\frac{1}{z_i}{\rm Res}_{z=z_i}I_{\rm term}(z)
    +{1\over 2\pi i}\int_{\cal C} \frac{dz}{z}I_{\rm term}(z).~~~\label{gen-cano-split}
\end{equation}
We can see that the right hand side gives a splitting of input
integrand. We want to show such a splitting, is nothing, but the
canonical splitting in unitarity cut method (\ref{cano-split}).

Now we calculate residues of poles for finite $z_i$. Since $G(\la)$
is pure holomorphic, we omit it during our derivation and put it
back only at the end. The BCFW deformation is given by\footnote{The
deformation null momenta $q$ can have two choices:
$\ket{\ell_1}\bket{\ell_2}$ or $\ket{\ell_2}\bket{\ell_1}$. These
two choices are equivalent to each other. Using $\ell_2=\ell_1+K$,
we have $\bket{\ell_2}\sim |K\ket{\ell_1}$.}
 $\begin{cases}
    \ket{\lambda}\to\ket{\lambda}\\ |\tl] \to |\tl]- z |K|\la\rangle
 \end{cases}$. After substituting the $|\tl]$ in the above equation, we  have
 \begin{equation}
    -\frac{1}{z_i}{\rm Res}_{z=z_i}I_{\rm term}(z)=-{1\over z_i}{\rm Res}_{z=z_i}\frac{\prod_{j=1}^{n+k}([a_j|\tl]-z[a_j|K|\la\rangle)}{\gb{\lambda|K|\tl}^{n+2}\prod_{i=1}^k(\gb{\lambda|Q_i|\tl}-z\braket{\lambda| Q_iK|\lambda})}
 \end{equation}
For a certain propagator $D_i$ on shell, we have
$z_i=\frac{\gb{\lambda|Q_i|\tl}}{\braket{\lambda|Q_iK|\lambda}}$ and
then we have
\begin{equation}
\begin{aligned}
&-\frac{1}{z_i}{\rm Res}_{z=z_i}I_{\rm term}(z)
=
\frac{1}{\gb{\lambda|Q_i\tl}}\frac{\prod_{j=1}^{n+k}([a_j|\tl]-\frac{\gb{\lambda|Q_i|\tl}}{\braket{\lambda|Q_iK|\lambda}}[a_j|K|\la\rangle)}{\gb{\lambda|K|\tl}^{n+2}\prod_{r\neq i}^k(\gb{\lambda|Q_r|\tl}-\frac{\gb{\lambda|Q_i|\tl}}{\braket{\lambda|Q_iK|\lambda}}\braket{\lambda|Q_rK|\lambda})}\\
=&\frac{1}{\gb{\lambda|Q_i\tl}\gb{\lambda|K|\tl}^{n+2}}\frac{\prod_{j=1}^{n+k}([a_j|\tl]\braket{\lambda|Q_iK|\lambda}-[a_j|K|\la\rangle\gb{\lambda|Q_i|\tl})}{\braket{\lambda|Q_iK|\lambda}^{n+1}\prod_{r\neq i}^k(\gb{\lambda|Q_r|\tl}\braket{\lambda|Q_iK|\lambda}-\gb{\lambda|Q_i|\tl}\braket{\lambda|Q_rK|\lambda})}
\end{aligned}
\end{equation}
Using a generalized version of Schouten identity \bea
\left|\tl\right]\vev{\la|Q_iK|\la}-|K|\la\rangle\gb{\la|Q_i|\tl}+|Q_i|\la\rangle\gb{\la|K|\tl}=0,\\
\left[\tl|Q_r|\la\right\rangle\langle\la|Q_i|-\left[\tl|Q_i|\la\right\rangle\langle\la|Q_r|+\vev{\la|Q_rQ_i|\la}[\tl|=0,
\eea the residue term will represent as
\begin{equation}
\begin{aligned}
&{1\over\gb{\lambda|Q_i|\tl}\gb{\lambda|K|\tl}^{n+2}}\frac{\prod_{j=1}^{n+k}(-\tgb{a_j|Q_i|\lambda}\gb{\lambda|K|\tl})}{(-1)^{n+1}\braket{\lambda |K Q_i|\lambda}^{n+1}\prod_{r\neq i}^{r=k}(-\braket{\lambda|Q_rQ_i|\lambda}\gb{\lambda|K|\tl})}
\\=&\frac{1}{\gb{\lambda|K|\tl}\gb{\lambda|Q_i|\tl}} \frac{\prod_{j=1}^{n+k} \tgb{a_j|Q_i|\lambda}}{\braket{\lambda |K Q_i|\lambda}^{n+1}\prod_{r\neq i}^{r=k}\braket{\lambda |Q_r Q_i|\lambda}}
\end{aligned}
\end{equation}
Putting it back to (\ref{gen-cano-split}), and comparing with
(\ref{cano-split}), we see that the first line of (\ref{cano-split})
is nothing, but the part ${1\over 2\pi i}\int_{\cal C}
\frac{dz}{z}I_{\rm term}(z)$, thus (\ref{gen-bubble-ceoff}) is
nothing, but taking residues of the first line of
(\ref{cano-split}). Thus we have shown the equivalence of getting
bubble coefficients in these two methods.

\section*{Acknowledgement}

We would like to thank Bo Feng for suggesting us  this project and
giving us many instructions along the way. Yang An is grateful to
his tutor Mingxing Luo. Z.Huang appreciates his tutor Ellis Ye Yuan
for his guidance. Yi Li is supported by  Chinese NSF funding under
contract No.11575156.
\bibliographystyle{unsrt}
\bibliography{note}

\end{document}